\documentclass[11pt,twoside]{article}

\usepackage{asp2006}
\usepackage{epsf}
\usepackage{psfig}
\usepackage{graphicx}
\usepackage{lscape}
\usepackage{amsbsy}
\usepackage{amssymb}

\begin{document}

\title{Turbulent magnetic fields in the quiet Sun: \\A search for cyclic variations}

\author{L.\,Kleint\altaffilmark{1,2}, S.\,V.\,Berdyugina\altaffilmark{3}, A.\,I.\,Shapiro\altaffilmark{4}, and M.\,Bianda\altaffilmark{2}}   

\altaffiltext{1}{Institute for Astronomy, ETH Z\"{u}rich, CH-8093 Z\"{u}rich, Switzerland}
\altaffiltext{2}{Istituto Ricerche Solari Locarno (IRSOL), CH-6605 Locarno Monti, Switzerland}
\altaffiltext{3}{Kiepenheuer-Institut f\"{u}r Sonnenphysik, Sch\"{o}neckstrasse 6, D-79104 Freiburg, Germany}
\altaffiltext{4}{Physikalisch-Meteorologisches Observatorium Davos, World Radiation Center, Dorfstrasse 33,  CH - 7260 Davos Dorf, Switzerland}

\begin{abstract} 
Turbulent magnetic fields fill most of the volume of the solar atmosphere. However, their spatial and temporal variations are still unknown. Since 2007, during the current solar minimum, we are periodically monitoring several wavelength regions in the solar spectrum to search for variations of the turbulent magnetic field in the quiet Sun. These fields, which are below the resolution limit, can be detected via the Hanle effect which influences the scattering polarization signatures ($Q/I$) in the presence of magnetic fields. 
We present a description of our program and first results showing that such a synoptic program is complementary to the daily SOHO magnetograms for monitoring small-scale magnetic fields.
\end{abstract}

\section{Introduction} 
 
Spectropolarimetry is a powerful tool for diagnosing magnetic properties of the Sun. Circular polarization, characterized by Stokes $V/I$, is produced by line-of-sight magnetic fields and is a basis of solar magnetograms. Linear polarization, characterized by Stokes $Q/I$ and $U/I$, is produced by transverse magnetic fields or through coherent scattering. A known disadvantage of magnetograms is their dependence on spatial resolution. For instance, let us imagine a resolution element that contains a line-of-sight magnetic field pointing towards us and another of equal strength pointing away from us. Their respective Stokes $V/I$ signals will be of opposite sign and, if they are at the same Doppler shift, their sum is zero. Hence, we observe a grey part in the magnetogram even though there may be strong fields below our resolution limit. This cancellation effect limits the use of magnetograms for small-scale, turbulent fields, and another technique has to be used for their detection.

\begin{figure*}[!ht]
 \begin{center}
 \plotone{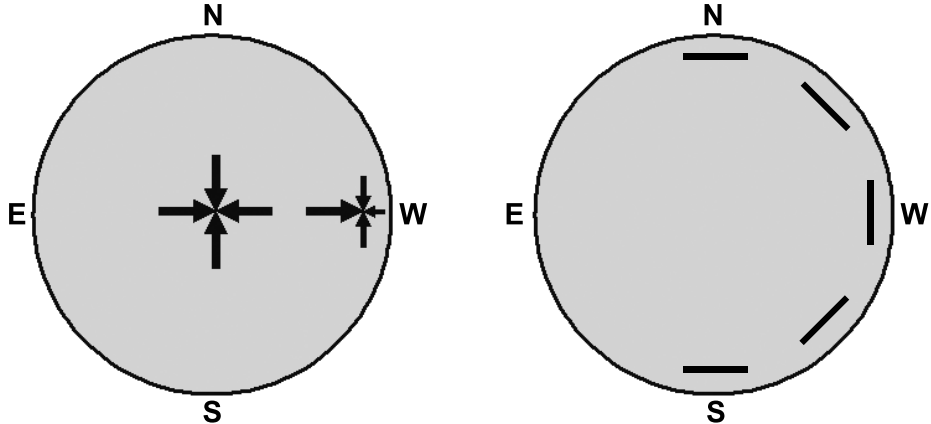} 
 \end{center}
\caption{ 
{\it Left}: Scattering geometry showing an equal amount of photons being scattered at disk center from every direction and thus no resulting polarization. At the limb, however, a radiation anisotropy results in a net polarization parallel to the nearest solar limb. 
{\it Right}: For our synoptic program, the spectrograph slit is positioned at five different heliographic angles, always at an equal distance to the solar limb in order to search for spatial variations of the turbulent fields (drawing not to scale).}
\label{scatt}
\end{figure*}
%
Coherent scattering processes on the Sun produce a linearly polarized spectrum, like scattering in the Earth's atmosphere produces polarization of the blue sky. The principle can be seen in Fig.~\ref{scatt}. Scattering polarization only arises if the spatial symmetry is broken by an anisotropic incident radiation field (the arrows). At the disk center, an equal amount of photons are scattered from every direction relative to the line-of-sight, and with no preferred direction there is no resulting polarization. However, at the solar limb the situation is different. The anisotropy introduces a preferred direction and a net linear polarization perpendicular to this direction, i.e., parallel to the next solar limb. On the Sun, the anisotropy is caused by the temperature gradient, which manifests itself in the limb darkening. The net polarization is produced in the continuum as well as in atomic and molecular lines which can depolarize the continuum or add on top of it. The linearly polarized $Q/I$ spectrum, with $Q/I$ being defined to be parallel to the next solar limb, is called the second solar spectrum \citep{ivanov91, stenflokeller97}. It does not resemble Stokes $I$ and cannot be easily deduced. 

Magnetic fields modify scattering polarization via the Hanle effect. Depending on the properties of atomic or molecular transitions, magnetic fields in the range from a fraction of Gauss to kilo Gauss can produce visible depolarization in $Q/I$ or rotate the plane of polarization and induce a signal in $U/I$. Independently of their spatial orientation, magnetic fields modify the $Q/I$ spectrum in the same way. Thus, in contrast to the Zeeman effect, the absence of cancellation enables us to detect fields below the resolution limit.

Our synoptic program monitors turbulent magnetic fields, their spatial distribution and variation with the solar cycle since 2007, starting before the appearance of the first sunspot of the cycle 24. This is the first systematic study of turbulent magnetic fields which can lead to a better understanding of small-scale fields governing most of the solar photosphere.

\section{The synoptic program}

Our synoptic program consists of monthly measurements of Stokes $IQV$ in three wavelength regions (5141 \AA, 5206 \AA \ and 4383 \AA) and at five position angles (N, NW, W, SW, S) around the solar limb at a distance of 5\arcsec \ (heliocentric angle $\mu=\cos\theta=0.1$) to the limb. The observations are carried out at the Istituto Ricerche Solari Locarno (IRSOL, Switzerland) using the 45 cm Gregory-Coud\'e telescope, the high-resolution spectrograph and the ZIMPOL polarimetry system \citep{gandorfer04}. Simultaneous recording of Stokes $V/I$ enables us to verify that we are targeting the quiet Sun. Recording these 15 measurements including setup, calibrations and positioning takes about two days.

Figure~\ref{scatt} shows the geometry of the observations with the five different slit positions (drawing not to scale). To achieve the desired signal-to-noise ratio, the signal is averaged over the 180\arcsec \ long slit. A more detailed description of the observing procedure and data reduction steps is presented in \citet{kleint10}. 

\begin{figure*}[!ht]
\plotfiddle{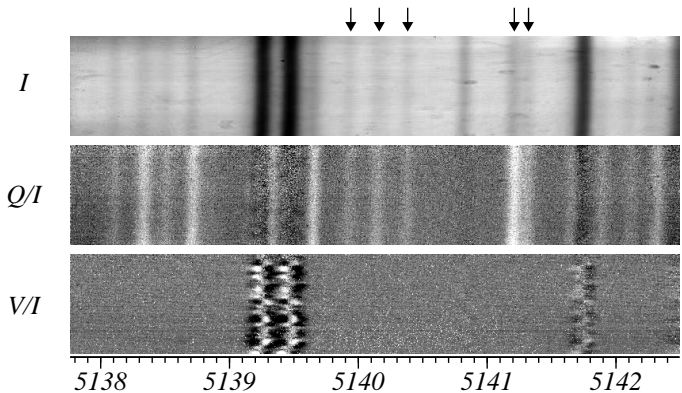}{4.5cm}{0}{90}{90}{-180}{-330}
\caption{Synoptic observation, taken on November 17, 2008, 5\arcsec \ from the south limb. The arrows denote the C$_2$ lines used for the analysis.} \label{synobs}
\end{figure*}

An example synoptic observation of the photospheric C$_2$ 5141\,\AA\ region can be seen in Fig.~\ref{synobs}. Stokes $I$ shows several strong Fe lines and some very weak C$_2$ lines. The C$_2$ lines of interest are marked with arrows in the figure. These are the resolved R-triplet (R$_1$ 5139.93\,\AA, R$_2$ 5140.14\,\AA, and R$_3$ 5140.38\,\AA), and the partly resolved P-triplet (P$_1$ 5141.21\,\AA, P$_2$ 5141.19\,\AA, and P$_3$ 5141.31\,\AA). These lines constitute an ideal set for measuring solar turbulent magnetic fields using the differential Hanle effect as described by \citet{sveta2004}. The linearly polarized $Q/I$ spectrum in this region ranges from -0.08 \% to 0.11 \% with the continuum fixed at zero. While the strong iron lines depolarize the spectrum, the C$_2$ lines are prominently visible, increasing the overall polarization. The $V/I$ image shows small-scale magnetic fields with signals ranging from -0.12 \% to 0.14 \%. These small fields cannot be avoided even at the most quiet places on the Sun and because of their size and their small polarization they would appear grey in SOHO magnetograms. Note that no $V/I$ signal is visible in any of the molecular lines due to their small Land\'e factors.

\section{Differential Hanle effect}

The influence of the Hanle effect on various spectral lines depends on their intrinsic properties. For instance, a critical magnetic field needed to depolarize a line to the saturation level depends on the upper level lifetime $t_{\mathrm{life}}$ and its Land\'e factor $g_L$ \citep[e.g.,][]{trujillospw3}
\begin{equation}
B_H \approx 1.137 \cdot 10^{-7} / (t_{\mathrm{life}} \cdot g_L)
\end{equation}
If several lines with different lifetimes and Land\'e factors but a similar formation height are observed within one spectral region, one can eliminate most sources of uncertainties by comparing directly polarization amplitudes of these lines, i.e., using a differential approach. Through these amplitude ratios a unique field strength can be found, independently of the filling factor or any model atmosphere. It is these qualities that make molecular lines a sensitive tool for employing the differential Hanle effect \citep{sveta2002,sveta2004}. We have therefore chosen the C$_2$ 5141\,\AA\  molecular region to be one of the primary targets for this study because of the six suitable lines with a large range of Land\'e factors and similar formation heights \citep{sveta2004}. The region around 5206 \AA \ could in principle also be interpreted differentially using the two chromospheric Cr~{\small I} lines. However, their formation heights may not be equal and further investigation is required to determine a reliable magnetic field strength. 

To classify the variations in the C$_2$ lines we calculate three amplitude ratios: R$_1$/R$_2$, R$_3$/R$_2$ and R$_2$/P where P denotes the total amplitude of P$_1$ and P$_2$. Variations of these ratios can directly be linked to fluctuations of the magnetic field, and variations even lower than a few Gauss could be seen (modeling of these ratios is presented by \citet{kleint10}).

\begin{figure*}[!ht]
\plotfiddle{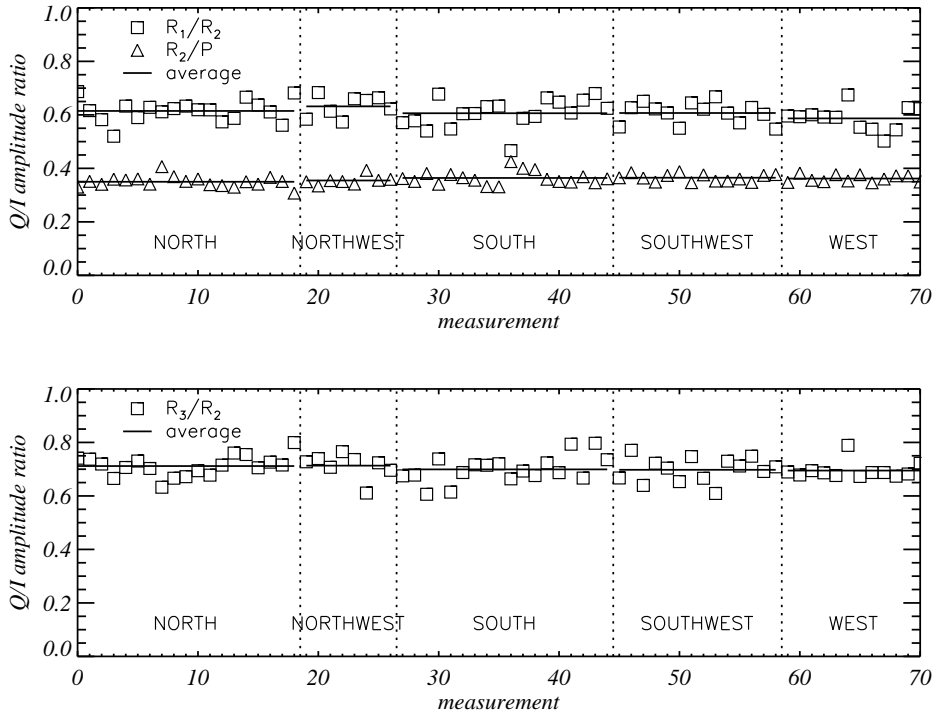}{9cm}{0}{78}{78}{-200}{-10}
\caption{Amplitude ratios of the synoptic measurements and their averages per position angle (solid lines). No variation can be seen for the past two years with the exception of one measurement (South, top panel) showing a deviation of more than 3$\sigma$ in R$_1$/R$_2$ and R$_2$/P.}\label{amprat}
\end{figure*}

\section{Results}

As of today our measurement series comprises of more than 70 measurements for the C$_2$ lines excluding all noisy measurements and measurements with limb distance outside the limit of 0.07 $< \mu <$ 0.13. 
Their amplitude ratios are plotted in Fig.~\ref{amprat}. The top panel shows R$_1$/R$_2$ and R$_2$/P with the solid lines denoting the averages per position angle. The lower panel shows the same for R$_3$/R$_2$. Within the error bars, the averages are the same for all position angles. Only one measurement in the south shows a deviation of more than 3$\sigma$ with respect to the average, showing a very similar amplitude ratio as a single measurement obtained at a solar maximum \citep{gandorferatlas}. We cannot conclude any variation from this single measurement but it is possible that the dispersion may get larger during a solar maximum or that the field strength would change. Further measurements are required during an increased solar activity to investigate this behavior.

\begin{figure*}[!ht]
\plotfiddle{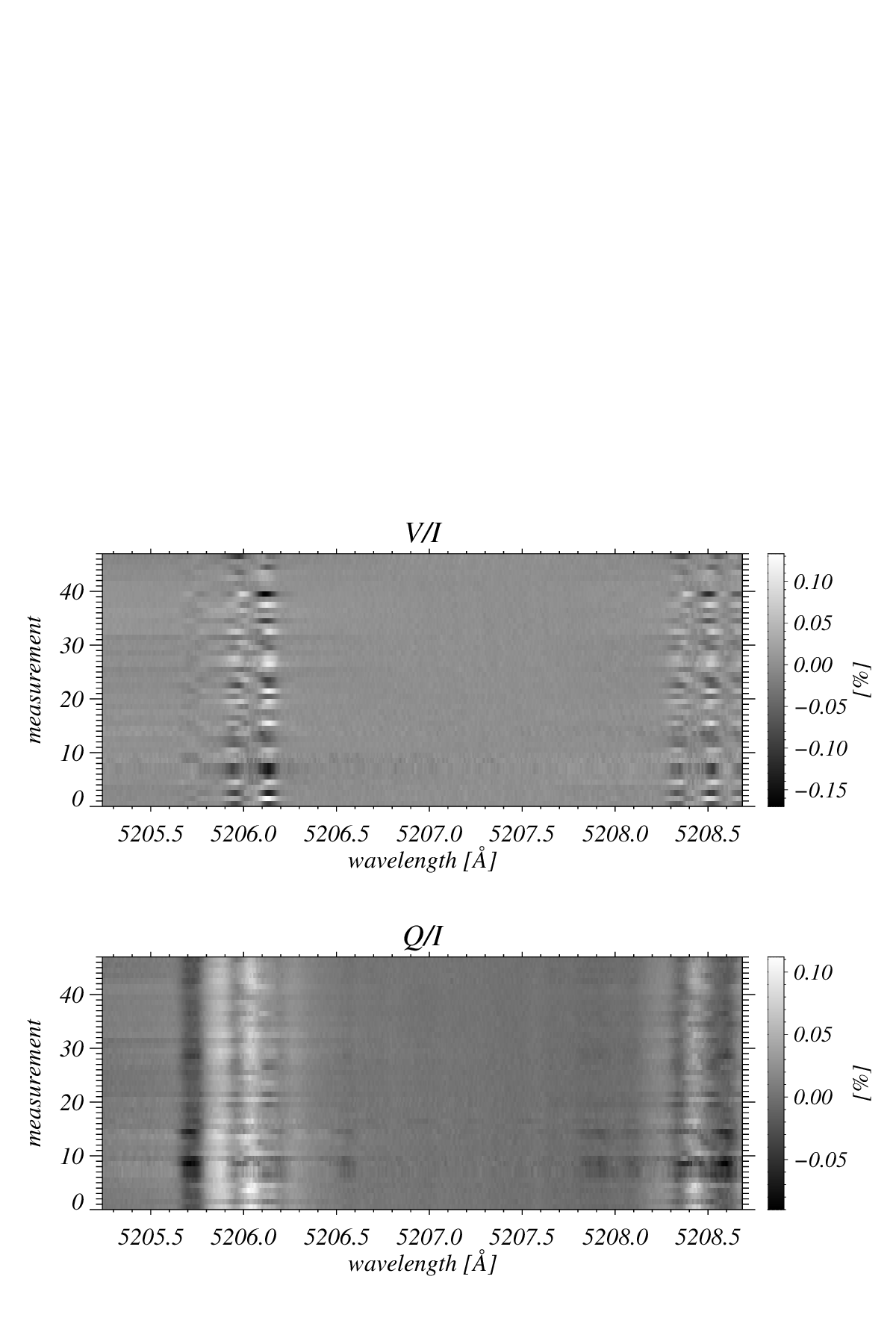}{9.0cm}{0}{57}{57}{-200}{-110}
\caption{Stokes $V/I$ and $Q/I$ of all synoptic measurements of the 5206 \AA \ region. Measurement 0 was taken in 2007 while  recent measurements were taken in 2009. Larger variations are visible in $Q/I$ than for the molecular lines which could be because of the higher formation height of the Cr~{\small I} lines.}\label{crvar}
\end{figure*}
%
Figure~\ref{crvar} shows all measurements of Cr~{\small I} with the vertical axis denoting the time. Because the measurements were not always taken with the exact same dispersion and grating angle which results in a wavelength shift, they were interpolated to a common wavelength scale. The separation into position angles was omitted because the sample does not contain a sufficient number of measurements to determine statistical properties. The top panel shows $V/I$ which again is significantly below 1\% as the targets were quiet Sun regions. The lower panel shows $Q/I$. It is well visible that the variations appear larger than for the photospheric C$_2$ lines. It should be noted however that the limb distance cannot be determined as precisely as for the C$_2$ measurements. Therefore, a lower overall $Q/I$ polarization may be because of a larger limb distance. This uncertainty however, does not cause any differential effects, for example the enhancement of the polarization in the 5206.05 \AA \ line around measurement 4 or the stronger depolarization in the 5205.7 \AA \ line.

\section{Conclusions}
The turbulent magnetic field as measured with the C$_2$ lines has not shown any variation during this current solar minimum. However, the average level of the polarization amplitudes is different from the single measurement obtained at the solar maximum. The chromospheric Cr~{\small I} lines at 5206 \AA \ show more variations, independent of position angle. More measurements are needed for their statistical analysis, and a model has to be developed for a quantitative interpretation. Our synoptic program represents a complementary way of measuring small-scale fields in addition to, e.g., polar field measurements. It is important to know how the turbulent field varies with time, also for different activity minima and maxima.

\acknowledgements 
We thank Dominique Fluri for valuable comments and Daniel Gisler, Peter Steiner and Renzo Ramelli for technical support. This work has been funded by the SNSF, grants 200020-117821 and 200020-127329. SVB acknowledges the \mbox{EURYI} (European Young Investigator) Award provi\-ded by the ESF (see www.esf.org/euryi) and the SNSF grant PE002-104552.

\end{document}